\documentclass[aps,prd,reprint,showpacs,floatfix,superscriptaddress]{revtex4-1}

\usepackage{graphicx}
\usepackage{dcolumn}
\usepackage{amssymb}
\usepackage{amsmath}
\usepackage{amsfonts}
\usepackage{hyperref}
\usepackage{breakurl}

\def\scr{$\tilde{\tau}_{\text{CR}}$}
\def\stau{$\tilde{\tau}_1$}

\def\lsp{$\chi$}

\begin{document}

\title{New gamma ray signal from gravitationally boosted neutralinos at the galactic center}

\author{M.~Cannoni}
\email{mirco.cannoni@dfa.uhu.es}

\author{M.~E.~G\'omez}
\email{mario.gomez@dfa.uhu.es}
\affiliation{Departamento de F\'isica Aplicada, Facultad de Ciencias Experimentales, 
Universidad de Huelva, 21071 Huelva, Spain}

\author{M.~A. P\'erez-Garc\'ia}
\email{mperezga@usal.es}
\affiliation{Departamento de F\'{i}sica Fundamental and IUFFyM, Universidad de Salamanca, Plaza de la Merced s/n 
37008, Salamanca, Spain}

\author{J.~D.~Vergados}
\email{vergados@uoi.gr}
\affiliation{Theoretical Physics Division, University of Ioannina, Ioannina, Gr 451 10, Greece}

%\date{25/4/2012}

\begin{abstract}
We discuss on the possibility that colliding dark matter particles in the form of neutralinos  
may be gravitationally boosted near the super-massive black hole at the galactic center
so that  they can have enough collision energy to annihilate into a stau pair. 
Since in some phenomenologically favored  supersymmetric models the mass splitting between
the neutralino and the lightest stau, one of 
the two scalar superpartners of the tau lepton, is a few GeVs, this channel may be 
allowed. In addition, staus can only decay into a tau lepton and another neutralino. We calculate 
the gamma-ray spectrum and flux generated by the tau pair discussing the observability of the obtained features.
\end{abstract}


\maketitle

Dark matter (DM) accounts for more than 80$\%$ of the mass of the Universe but  
its nature is still one of the open problems in Physics. 
In a widely accepted theoretical scenario, DM
is formed by a weakly interacting massive particle (WIMP) that has  been in 
thermal equilibrium with Standard Model (SM) matter in the early Universe, leaving, after decoupling, 
the DM relic density as inferred by WMAP~\cite{WMAP}. 
In this light, supersymmetric (SUSY) extensions of the SM provide a 
natural WIMP candidate. In the minimal supersymmetric standard model (MSSM), 
R-parity conservation assures that if the lightest supersymmetric particle 
is the lightest of the four neutralino states--indicated as 
$\chi$ in the following--, this particle is absolutely stable.
In a phenomenologically favored scenario of the  constrained MSSM (CMSSM), the  stau 
coannihilation region (\scr)~\cite{stauco}, the lightest stau, $\tilde{\tau}_1$, one 
of the scalar super-partners of the tau lepton, is close in mass to the neutralino. 
In the \scr~ parameter space the cross 
section for non-relativistic annihilation into fermions of the SM, $\chi \chi \to f\bar{f}$, is 
typically small and results in a too large relic density. However, including 
the so-called coannihilation processes~\cite{Griest}, as 
for example  $\chi \tilde{\tau}_1$, $\tilde{\tau}_1 \tilde{\tau}_1$ collisions, 
when the  mass splittings of the involved particles are small,
one can efficiently enhance the thermally  averaged cross section $\langle \sigma v\rangle$, 
and, consequently, diminish the relic density to the measured value.

The standard cosmological model predict that 
non-relativistic cold DM particles ($v/c \sim 10^{-3}$) 
cluster into halos~\cite{halo} that contain baryonic matter. 
Since DM  in the halo follows a certain 
mass distribution, the two-body annihilation processes can happen at a rate that is proportional to the 
DM mass density  squared. Therefore, the highest chances to detect an observable indirect signal of 
their existence are attained in a region with high DM density, in particular, in the galactic 
center (GC).
Among the various signatures from DM annihilation,
gamma-ray signals have received much attention. 
A continuum spectrum of secondary photons may arise from hadronization and decay of the annihilation 
products~\cite{gammas} and from radiation from final state charged particles~\cite{IB}. 
Direct annihilation into photons is also possible but only at loop level~\cite{lines}.

The gravitational potential in the GC is dominated by a super-massive
black hole (BH) with mass $M_{\text{BH}}=4\times 10^6 M_\odot$
and Schwarzschild radius $R_{\text{S}}=2GM_{\text{BH}}/c^2=4\times 10^{-7}$ pc~\cite{SMBH}. 
Recently, the idea that a BH can act as a particle accelerator has been 
proposed~\cite{Banados1}.  
The highest center of mass frame (CMF) energies
are obtained when the colliding particles approach the 
horizon on falling geodesics with opposite angular momentum per unit mass $L/m_{\chi}\leq 
L_{\text{c}}=4GM_{\text{BH}}/c$,
in the case of the Schwarzschild metric.
The maximum possible value is $\sqrt{s}=2\sqrt{5}m_{\chi}$~\cite{Baushev,Banados1} 
for a non-rotating BH, while it can be arbitrarily large for the Kerr BH~\cite{Banados1}.
In principle, due to this general relativity effect, new annihilation channels into heavier states, 
kinematically forbidden for non relativistic particles, could be accessible.
Additionally, a realistic calculation of an  indirect DM signal in this scenario would also be determined by the 
particle escape function at  distances close to the BH. For the horizon proximity this has been calculated in 
\cite{Banados2} under the restrictive assumptions of annihilation into two massless particles with isotropic angular 
distribution.

In this work we show that if DM is formed by neutralinos with the characteristics of the \scr,
a new dominant annihilation channel may 
be opened already for sub-relativistic neutralinos~\cite{Amin} boosted in the inner regions of the GC.

Near the BH the DM density is described by a power-law $\rho(r)\propto r^{-\gamma}$, as we will discuss later.
From the  Newtonian approximation given by the Jeans equation, the root mean squared velocity 
is $v(r) \approx (G M_{\text{BH}}/r)^{1/2}$~\cite{BT,Gnedin,MerrittLet}, or in terms of the 
Schwarzschild radius, $v(r)/c \approx (R_{\text{S}}/2r)^{1/2}$. 
Since a Keplerian orbit with $L_{c}$ would cross the horizon if the pericenter 
distance is less than 
$r_{\text{min}}=4R_{\text{S}}$, hence we will consider safely $r>4R_{\text{S}}$. In this way, for example, at $r=10 
R_{\text{S}}$ we have $v/c\simeq 1/\sqrt{20}\simeq 0.22$.
Requiring that the neutralino has the relic density inferred by WMAP, the relative mass splitting with the \stau, 
$\delta =(m_{\tilde{\tau}}-m_{\chi}) / m_{\chi}$, is typically less than $5\%$.
In the CMF the energy threshold for stau pair production is
$\sqrt{s}=2E_{\chi} \geq 2m_{\tilde{\tau}}$, that implies $v/c \geq [{1-{1}/(1+{\delta})^2}]^{1/2}$. With 
$\delta=2\%$, $v/c \geq 0.197 $:
there exists thus  a range of radii  where the kinetic energy is high enough to reach the threshold of the process
$\chi \chi \to \tilde{\tau}^{-} \tilde{\tau}^{+}$.
The maximum radius is given by $r_{\text{max}}= 1/2[1-(m_{\tilde{\chi}}/m_{\tilde{\tau}})^2]R_{\text{S}}$.

\begin{figure}[t!]
\includegraphics*[scale=0.45]{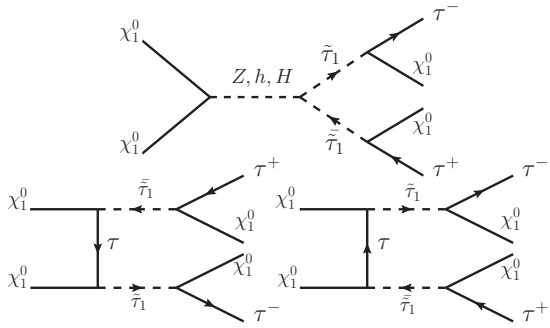}
\caption{Diagrams for stau pair production and decay in neutralino annihilation.}
\label{fig:1}
\end{figure}

If the mass splitting $\Delta m=m_{\tilde{\tau}}-m_\chi$ is larger than the tau mass, 
$m_\tau =1.777 $ GeV, the staus can only decay into the two body final 
state $\chi\tau$, see diagrams in Fig.~\ref{fig:1}.
The neutralino is bino-like thus the vertices's \lsp--\lsp--$Z$ and 
\lsp--\lsp--($h,H$) are suppressed, while the vertex \lsp--\stau--$\tau$
is not suppressed by mixing. In fact it is proportional to $Z_{11} U_{12}$, the product of the 
relevant neutralino and stau mixing matrix elements that are both close to one.
The dominant diagrams in Fig.~\ref{fig:1} are thus the ones with $t$, $u$ channel
exchange of the tau.
At energies near the threshold the produced staus are slow thus the propagator
$1/(p^2_{\chi} -p_{\tilde{\tau}})^2 -m^2_{\tau} =1/(m^2_{\tilde{\tau}} +
m^2_{\chi} -2E_{\chi} E_{\tilde{\tau}} + 2\mathbf{p}_{\tilde{\tau}}\cdot 
\mathbf{p}_{{\chi}} -m^2_{\tau}    )$
is approximately $1/[(m_{\tilde{\tau}}-m_{\chi})^2-m^2_\tau]$. The cross section, 
proportional to the square of this quantity is, thus, enhanced for mass splittings
approaching the tau mass. 
%
%%%%%%%%%%%%%%%%%%%%%%%%%%%%%%%
%%%%%%%%%%%%%%%%%%%%%%%%%%%%%%%
%
 
\begin{figure}[t!]
\includegraphics*[scale=0.52]{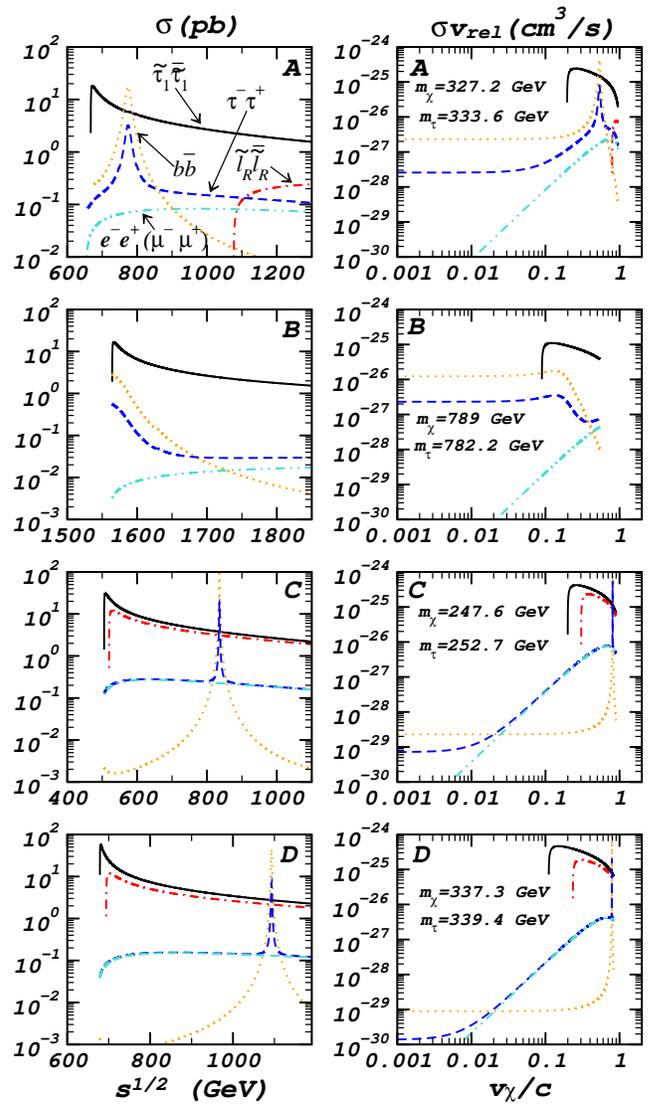} 
\caption{Annihilation cross sections in picobarn as a function of the CMF energy (left panels) and annihilation 
cross section times the relative velocity in cm$^3$/s as a function of the colliding neutralino velocity (right 
panels). The CMSSM points are specified in Table~\ref{tab:1} and the annihilation channels are labeled
in the upper left panel.}
\label{fig:2}
\end{figure}

We illustrate the above features in Fig.~\ref{fig:2},
where we show the relevant cross sections as a function of $\sqrt{s}$, left panels,
for four points of the \scr ~that are allowed by present phenomenological
constraints. The values of the universal scalar mass $m_0$, gaugino mass $m_{1/2}$, trilinear
scalar coupling $A_0$ and the ratio of the two Higgs expectation values $\tan\beta$ that define
the CMSSM parameter space are given in Table~\ref{tab:1}.
\begin{table*}[htbp!]
\caption{CMSSM points used in his work.
The sign of $\mu$ is positive. The neutralino and stau masses are also reported.} 
\begin{ruledtabular}
\begin{tabular}{ c  c  c  c  c  c c }
%\hline
 & $m_{0}$ (GeV) & $m_{1/2}$ (GeV) & $A_{0}$ (GeV) & $\tan\beta$ & $m_{\tilde{\chi}}$ (GeV)
 & $ m_{\tilde{\tau}_1}$ (GeV)\\
\hline
%\hline
A & 452 & 780  & 1110 & 41 & 327.2  & 333.6 \\
%\hline
B & 858 & 1780  & 0  & 45  & 789.0   &  782.2   \\
%\hline
C & 122 & 600   & 0  & 10 & 247.6  &  252.7       \\
%\hline
D & 166 & 805   & 0  & 10 & 337.3   &  339.4       \\
%hline
%\hline
\end{tabular}
\end{ruledtabular}
\label{tab:1}
\end{table*}
The numerical computation was done using the interfaced codes 
$\textsf{MicrOMEGAs}$~\cite{Micromegas}, $\textsf{CalcHEP}$~\cite{Calchep} 
and $\textsf{SOFTSUSY}$~\cite{Softsusy}. 
The point A is similar to best-fit point found in ~\cite{bestfitCMSSM},
that predict a light Higgs around 119 GeV. 
Next we take three cases with  $A_0 =0$, as usual for setting upper limits with LHC searches. 
In the point B the Higgs is slightly heavier. 
Points  C and D have low $\tan\beta$ and the Higgs is around 115 GeV.
In all cases, the cross section for stau pair-production clearly dominates
by one or two orders of magnitude the cross sections for annihilation 
into fermions except when $\sqrt{s}$ corresponds to the  heavy Higgs $A,H$ masses
where the $s$-channel propagators are resonant. 
%
%%%%%%%%%%%%%%%%%%%%%%%%%%%
%%%%%%%%%%%%%%%%%%%%%%%%%%%%
%
%
In the right panels in Fig.~\ref{fig:2} we show the annihilation cross section times the particle relative velocity, 
as a function the CMF velocity of colliding neutralinos; this is the quantity that enters in 
the calculation of indirect detection signals of the processes considered. 
Note that $\sigma v_{\text{rel}}$ for annihilation in staus, near the threshold,
is at least an order of magnitude bigger than the freeze-out value $3\times 10^{-26}$ cm$^3$/s
and that these values corresponds to  $v/c\sim 0.1-0.2$ that are just the ones that can be obtained 
with the gravitational boost discussed above. 
At low $\tan\beta$, cases C and D, the "right" selectron and smuon ($\tilde{\ell}_{\text{R}}$)  tend to become 
degenerate in mass with 
$\tilde{\tau}_1$ and the cross section for annihilation into pairs of these scalars is much larger than in 
the cases A and B. 
Although the masses of the particles in point B are much heavier than in the other cases, the mass splitting 
is around 3 GeV and $\sigma v$ is of the same magnitude.
The same effect can be seen comparing case D with C.

A possible signal of the opening of the new channel is given by
the gamma-rays produced by the tau pair.
The extension of the source is set by $r_{\text{max}}$. This is too small to 
be resolved by present telescopes, thus we treat it as point source at the GC 
at a distance from us of $D=8$ kpc. 
To evaluate the flux we first note that 
applying the small width approximation to the stau propagators, and
given that $BR(\tilde{\tau}^{\pm}_1 \to \tau^{\pm}\chi) =1$, we have 
$\sigma(\chi \chi \to \tau^{-} \tau^{+} \chi \chi)
\simeq \sigma(\chi \chi \to \tilde{\tau}_1 \overline{\tilde{\tau}}_1)
BR^2(\tilde{\tau}_1 \to \tau \chi)
\simeq
\sigma(\chi \chi \to \tilde{\tau}_1 \overline{\tilde{\tau}}_1)
\equiv \sigma_{\tilde{\tau}  \tilde{\tau}}.$
We can thus evaluate the differential photon flux as
\begin{equation}
\frac{d\Phi}{dE_\gamma}=\frac{R_{\text{S}}^3}{ D^2}
\int\limits_{r_{\text{min}}}^{r_{\text{max}}} dr  r^2
\sigma_{\tilde{\tau}  \tilde{\tau}}(r)
v_{\text{rel}}(r) \frac{\rho^2(r)}{m_{\chi}^2}
\frac{dN}{dE_\gamma}(r) .
\label{flux}
\end{equation}
In the integral we treat the distances in units of the Schwarzschild radius,
thus $r$ is dimensionless and a factor $R_{\text{S}}^3$ appears explicitly. 
We note some differences with the standard almost-static $\chi\chi\to\tau^+\tau^-$
annihilation: 
(i) there is no factorization into a particle physics and astrophysics factor because all the factors 
in the integrand depend on $r$ through the velocity dependence.
An integration over the CMF scattering angle is implied in $\sigma_{\tilde{\tau} \tilde{\tau}}(r)$
that is evaluated taking the exact spin averaged squared matrix elements from  $\textsf{CalcHEP}$; 
(ii) we do not divide by 2 because the final state necessarily contains two neutralinos; 
(iii) the taus are not monochromatic and the spectrum changes with the collision energy $\sqrt{s}$ and
ultimately with the distance, 
while in the static case the taus have an energy equal to the neutralino mass and 
the radiated photon spectrum is limited by $E_\gamma^{\text{max}} =E_\tau =m_{\chi}$. 

Before proceeding further we will discuss this last point.
The taus energy spectrum can be easily obtained by
applying a Lorentz transformation with parameters $\beta=({1-{4m^2_{\tilde{\tau}}}/{s}})^{1/2}$ and
$\gamma={\sqrt{s}}/{2m_{\tilde{\tau}}}$,
to the spectrum calculated in the rest frame of the stau. 
In this frame it has fixed energy and momentum,
$E_{\tau}^* =(m^2_{\tilde{\tau}}-m^2_{\chi} +m_{\tau}^2)/{2m_{\tilde{\tau}}}$
and $p_{\tau}^* =({E_{\tau}^*}^2 -m_{\tau}^2 )^{1/2}$.
The resulting energy distribution is flat and limited,
${dN(\tilde{\tau}_1 \to \chi \tau)}/{dE_{\tau}}=
{1}/{\Delta E}$,
$\Delta E =E_\tau^{\text{max}} - E_\tau^{\text{min}}$,
with 
$E_\tau^{\text{min}} =\gamma(E_{\tau}^*-\beta p_{\tau}^*)\leq
E_{\tau}\leq
E_\tau^{\text{max}}=\gamma(E_{\tau}^*+\beta p_{\tau}^*)$.
The number of photons with energy $E_\gamma$
produced by a tau with energy $E_\tau$
is given by $dN_\gamma /dx=1/2 f(x)$ with $x=E_\gamma / E_{\tau}$ 
and $f(x)=x^{-3/2}\exp[g(x)]+q\log[p(1-x)](x^2-2x+2)/x$.
This formula was obtained in Ref.~\cite{Cembranos}, to which we refer the reader 
for details, 
by fitting the photon yield from taus obtained with Monte Carlo simulations 
of the non relativistic process $\chi\chi\to\tau^+\tau^-$.
In this case the taus have equal energy, hence we use a factor $1/2$ for the yield
of one particle.
The gamma spectrum at distance $r$ is then obtained
by integrating over the tau energy distribution, 
\begin{equation}
\frac{dN}{dE_\gamma}(r) =\frac{1}{\Delta E(r)}
\int\limits_{E_\tau^{\text{min}}(r)}^{E_\tau^{\text{max}}(r) }
\frac{dE_\tau}{E_\tau}f\left(\frac{E_\gamma}{E_\tau}\right)
\theta(E_\tau -E_\gamma).
\label{eq:gammaspectrum}
\end{equation}
We have multiplied by 2 to obtain the yield of the pair.
The Heaviside function takes into account that for 
fixed $E_\gamma$ the integrand is zero if $E_\gamma > E_\tau$.
For this reason  the photon energy cut-off is 
$E_\tau^{\text{min}}$ for each $\sqrt{s}$.
The absolute cut off when integrating over $\sqrt{s}$ will be 
at $\gamma(r_{\text{max}})E_{\tau}^*$.
In fact as $r \to r_{\text{max}}$, $\beta \to 0$ and $E_\tau^{\text{min}}\to \gamma(r_{\text{max}})E_{\tau}^*$.
Note that $\gamma(r_{\text{max}})\sim 1$ and $E_{\tau}^* \lessapprox (m_{\tilde{\tau}}-m_{\chi}) $,
thus the cut off is indicative of the mass splitting between
the neutralino  and the stau.

The last ingredient that we need to evaluate in Eq.~(\ref{flux}) is the DM density profile. 
In~\cite{GondoloSilk} it was shown that the adiabatic growth of the BH at the 
center of the halo causes a steepening, called spike, of the initial halo profile toward the GC. 
Successive studies
~\cite{Bertone,MerrittReview,Ullio,Gnedin,MerrittLet,BertoneMerritt,MerrittCrest,Vasiliev1,Vasiliev2}  
showed that considering  physical effects such as scattering of DM particles off stars,
capture by the BH, self-annihilation and capture within stars 
during the evolution of the DM distribution, results in a shallower profile, 
$\rho_{\text{sp}}(r)\propto r^{-3/2}$.
For the SUSY models A-D, and generally for all the \scr ~parameter space, the spin independent and spin dependent 
elastic neutralino-nucleon 
cross sections are in the range 10$^{-11}$-10$^{-9}$ pb and 10$^{-8}$-10$^{-7}$ pb~\cite{CannoniES}, respectively,
thus the energy lost by elastic collisions with baryonic matter is not likely to be important.
At a certain distance from the GC the density reaches a value such that self-annihilation 
itself acts to stop further rising. 
It has been shown~\cite{Vasiliev1,Vasiliev2} that this latter effect do not set the density to a constant value, 
usually called annihilation plateau or core, but results in a mild spike (MS) with 
$\rho_{\text{ms}}(r)\propto r^{-1/2}$. 
In addition, the adiabatic compression of the gravitational potential caused by the baryons already in the bulge of 
the galaxy~\cite{Prada} should also be taken into account. 
%
%%%%%%%%%%%%%%%%%%%%%%%%%%%%
\begin{figure}[t!]
\includegraphics*[scale=0.5]{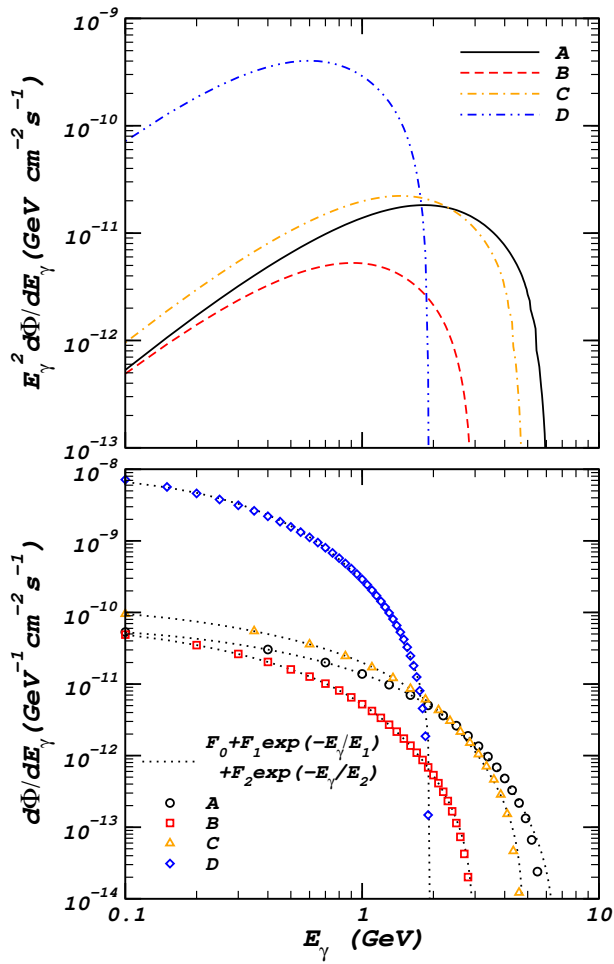} 
\caption{Top panel: differential flux multiplied by $E_\gamma^2$. 
Bottom panel: Differential flux (dots) and fit (dashed lines) with
$F(x)=F_0 + F_1 \exp(-E_\gamma /E_1) +F_2 \exp(-E_\gamma /E_2)$.
$F_0$, $F_{1,2}$, $E_{1,2}$ are fit parameters.}
\label{fig:4}
\end{figure}
%%%%%%%%%%%%%%%%%%%%%%%%%%%%%
%

We hence model the profile considering that at the radius 
$r_{\text{sp}}\approx 0.2 r_{\text{h}} $, with $r_{\text{h}}=1.67 $ pc 
the influence radius of the BH, DM density is given by a compressed Einasto profile, $\rho_\text{sp} =5\times 10^6$ 
GeV/cm$^3$ as in~\cite{Bertone}. 
From here the profile is given by $\rho(r)=\rho_{\text{sp}}({r}/{r_{\text{sp}}})^{-\gamma_\text{sp}}$,
$\gamma_\text{sp}=3/2$, up to the radius $r_{\text{a}}$ where the density reaches 
the value $\rho_{\text{a}}={m_{\chi}}/(\sigma v)_0 t_f$.
$(\sigma v)_0$ is the annihilation cross section  
and $t_f=10$ Gyr~\cite{Bertone,MerrittReview}
is the elapsed time since the formation of the spike. 
Finally, the inner MS is $\rho_{\text{a}}({r}/{r_{\text{a}}})^{-\gamma_\text{a}}$, $\gamma_\text{a}=1/2$,
up to the limit $4R_{\text{S}}$.
The radius $r_{\text{a}}$ is found by matching the two power-laws,
$r_{\text{a}} =r_{\text{sp}} (\rho_{\text{a}}/\rho_{\text{sp}})^{-1/\gamma_{\text{sp}}}$.
The values of $\rho_{\text{a}}$ are between $10^{11}-10^{12}$ GeV/cm$^3$ 
and $r_{\text{a}} $ are of the order $10^{-4}$ pc
for the CMSSM points A-D. In the cases C and D,  
$\sigma v$ in the non relativistic  limit is around $ 10^{-29}$ cm$^3$/s
that would result in $\rho_{\text{a}}$ a factor $10^2$
larger than in A and B. Anyway, as was already noted in~\cite{Amin}, the cross section for 
annihilation into leptons is strongly velocity dependent as can be seen in Fig.~\ref{fig:2}:
it rises rapidly reaching values around $10^{-27}$ cm$^3$/s at $v/c\sim 0.1-0.2$,
thus $\rho_{\text{a}}$ is of the same order as in A and B.

In the top panel of Fig.~\ref{fig:4} we show the differential flux multiplied
by $E_\gamma^2$ to exhibit the behavior at the highest energies near the cut off.
In the bottom panel, the differential photon flux is given by the dots.
We find that the spectral shape of the flux is well fitted with 
the sum of two exponentials, as shown by the dashed lines in the bottom panel.
The functional form is $F(x)=F_0 + F_1 \exp(-E_\gamma /E_1) +F_2 \exp(-E_\gamma /E_2)$
with $F_0$, $F_{1,2}$, $E_{1,2}$ fit parameters.

The peculiar characteristics of the signal are: (i) its origin is in
the innermost region around the BH where the DM distribution is given 
by the MS density; (ii)  the differential flux presents a nearly exponential shape with a hard cut-off
that is determined by the mass splitting between the neutralino and the stau; 
(iii) the signal shows up at energies below 10 GeV. 
Interestingly, in this few GeV energy region there are some unexplained excesses over the known 
backgrounds~\cite{fermi,hooper}.
However our 
predicted signal is too feeble to account for them. 
It is expected that after 5 years operation, the Fermi-LAT satellite reaches
sensitivities of $10^{-10}-0.5\times 10^{-11}$ 
photons cm$^{-2}$ s$^{-1}$ for energies between 0.5 GeV and 10 GeV~\cite{aldo}.
The proposed signal can be
one of the components observed by the collaboration. Furthermore, it
might be discriminated by the new proposed experiment Gamma-Light~\cite{aldo} that should
achieve a better energy and angular resolution than Fermi-LAT in the interval (10 MeV-1 GeV).

In summary, we have shown that if DM is formed by neutralinos as described in the stau coannihilation region
of the CMSSM,  stau pair production  may be the dominant annihilation channel in the innermost region 
of the GC near the BH. We 
have further shown that the gamma-ray spectrum produced by the $\tau\tau$ pair coming from the 
decay of the staus present peculiar features. 
This may further motivate the exploration of the GC by the Fermi-LAT satellite
to achieve a better understanding of backgrounds and to look for a possible new indirect signal of the presence of a 
DM component.

%\end{document}

\paragraph*{\textsl{Acknowledgements --}}
M.~C.~is a MultiDark fellow. Work supported by MultiDark under grant number CSD2009-00064 of the
Spanish MICINN Consolider-Ingenio 2010 Programme. Further support is provided by:
MICINN projects FPA2011-23781, 
FIS-2009-07238, MICINN-INFN(PG21)AIC-D-2011-0724, ESF-COMPSTAR and Junta de Andalucia under grant P07FQM02962.
M.~C. acknowledges the hospitality of the Fundamental Physics Department of University of Salamanca where part
of this work was developed. The authors acknowledge J.~Cembranos, G.~Gomez-Vargas, A.~Morselli, R. Lineros and 
M.~A.~Sanchez-Conde for useful discussions.

\end{document}